\documentclass[aps,pre,reprint,showpacs,superscriptaddress]{revtex4-1}

\usepackage{graphicx}
\usepackage{subfigure}
\usepackage{color}
\usepackage{amsmath}
\usepackage{lipsum}
\usepackage{amssymb}
\usepackage{epstopdf}
\usepackage[linktocpage,colorlinks=true,linkcolor=blue,citecolor=blue,breaklinks=true]{hyperref}
\usepackage{verbatim}
\usepackage{breakcites}
\usepackage[margin=0.95in]{geometry}
\usepackage{ulem}

\begin{document}


\title{Absence of Evidence for the Ultimate Regime in Two-Dimensional Rayleigh-B\'enard Convection}

\author{Charles R. Doering}
\affiliation{University of Michigan, Ann Arbor, Michigan 48109-1043, USA}
\author{Srikanth Toppaladoddi}
\affiliation{University of Oxford, Oxford OX2 6GG, United Kingdom}
\author{John S. Wettlaufer}
\affiliation{Yale University, New Haven, Connecticut 06520-8109, USA}
\affiliation{University of Oxford, Oxford OX2 6GG, United Kingdom}
\affiliation{Nordita, Royal Institute of Technology and Stockholm University, SE-10691 Stockholm, Sweden}
\maketitle

Zhu {\it et al.}~\cite{zhu2018} reported direct numerical simulations of turbulent thermal convection in two dimensions (2D) with planar no-slip isothermal walls over six decades of Rayleigh numbers $Ra \in [10^8, 10^{14}]$.
The authors reported scaling of the Nusselt number $Nu \sim Ra^{0.35}$ for the final four data points with $Ra \in [10^{13},10^{14}]$.
They also decomposed $Nu$ into contributions from ``plume-ejecting" ($Nu_e$) and ``plume-impacting" ($Nu_i$) regions of the spatial domain reporting $Nu_e \sim Ra^{0.38}$ for those four data points, interpreting this as evidence of a so-called `ultimate' regime of thermal convection characterized by {\it bulk} heat transport scaling $Nu \sim Ra^{1/2}$ modulo logarithmic corrections \cite{Kraichnan62,Chavanne97}.

Although hypotheses concerning the nature of boundary layers constitute one ingredient of this system, the essential characterization of the state of convection is the asymptotic $Nu$-$Ra$ relation \cite{LPK2001}.
Zhu {\it et al.}~\cite{zhu2018} drew an arbitrary line through the final four heat flux data with $Ra \in [10^{13},10^{14}]$ but when we performed an objective least-squares power law fit to these data \cite{DTW} we found $Nu =  0.035 \times Ra^{0.332}$ with an empirical exponent that is {\it indistinguishable} from $1/3$, the so-called `classical' scaling exponent \cite{priestley, malkus, howard}.
That said, under modern statistical standards a power law fit to experimental data indicates ``scale free" behavior only when observations extend over \textit{at least two decades}.
Failing this minimal criterion, scientific significance cannot be attributed to such power laws \cite{powerlaw}.
Importantly, however, a power law fit of the last eight data points for $Ra \in [2 \times 10^{12},5 \times 10^{14}]$, including two additional data points that Zhu {\it et al.} have since produced \cite{ZhuReply,remark}, yields $Ra^{0.337}$ scaling.  

Interestingly, the data from $Ra = 10^8$ to $10^{13}$ are extremely well described by extrapolation of a previous fit, $Nu = 0.138 \times Ra^{2/7}$, from high resolution simulations for $Ra \in [10^7, 10^{10}]$ \cite{charlie2009}.
Indeed, the objective least-squares power law fit of those nearly five decades of data for $Ra \in [10^8, 10^{13}]$ yields the scaling exponent 0.289, indistinguishable from $2/7$.

\vspace{0.10cm}
\begin{figure}[h]
\vspace{0.20cm}
\centering
\includegraphics[scale = 0.180]{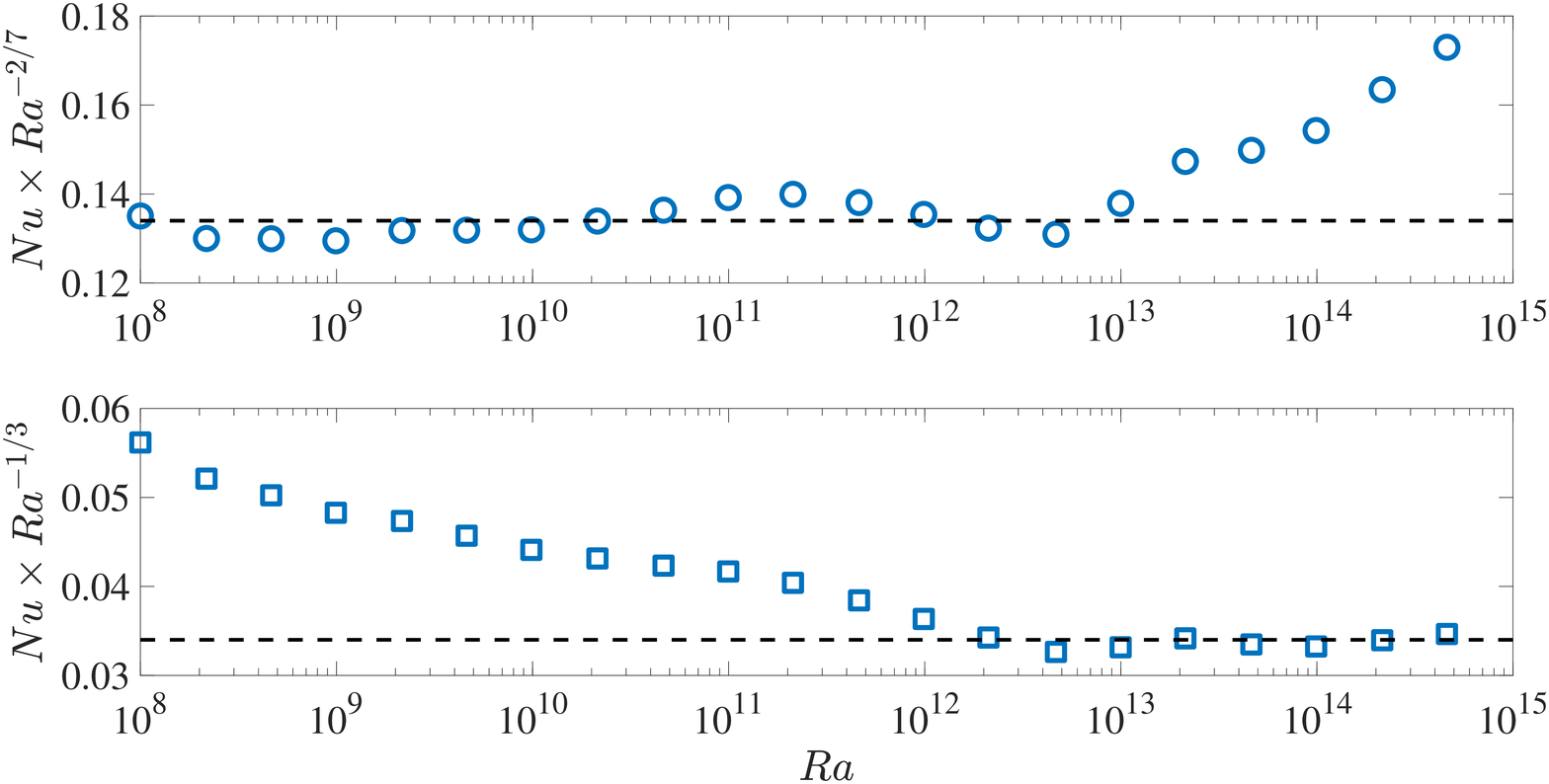}
\caption{Compensated plots of $Nu \times Ra^{-\beta}$ vs. $Ra$. Circles (top) are data of Zhu {\it et al.} --- from their Letter \cite{zhu2018} including two additional data points they subsequently produced \cite{ZhuReply} --- with $\beta = 2/7$ while squares (bottom) are the same data with $\beta = 1/3$.}
\label{fig:NuRa2}
\vspace{-0.6 cm}
\end{figure}

Compare Fig.~\ref{fig:NuRa2} here to Figs.~1 of  \cite{zhu2018} and \cite{ZhuReply,remark}.
Clear deviation from pure scaling combined with the range of $Ra$ and the small size of the data set over which the classical $1/3$ scaling appears precludes definitive extrapolation to asymptotically large $Ra$.
We note, however, that Fig.~\ref{fig:NuRa2} here shows that the classical exponent fits the last eight data spanning two full decades.
Moreover, the heat transport results reported by Zhu {\it et al.}~\cite{zhu2018,ZhuReply} are consistent with previous 3D simulations \cite{VC,SVL} and experiments~\cite{urban1,AGL_FIG.4,Sun} that suggest crossovers from $Nu \sim Ra^{2/7}$ to $Nu \sim Ra^{1/3}$ for various $Ra \in [10^9, 10^{11}]$.

To summarize, while Zhu {\it et al.}~did not report any consistent statistical analysis of their data, we have shown that when dividing their data into different segments and forcing a power law fit to the last decade as they proposed, the reduced data set is consistent with $Nu \sim Ra^{1/3}$ in quantitative agreement with the classical theories \cite{priestley, malkus, howard} and {\it not} the $Ra^{0.35}$ scaling reported in \cite{zhu2018}.
Inclusion of the additional two data produced to extend the high Raleigh number segment \cite{ZhuReply} does not alter this observation.
Furthermore, the remaining data are consistent with $Nu \sim Ra^{2/7}$ scaling in qualitative accord with other 2D \cite{charlie2009} and 3D results \cite{VC, SVL, urban1, AGL_FIG.4, Sun}.

Therefore the claim by Zhu {\it et al.} that their simulations suggests that 2D convective turbulence reaches an `ultimate' regime characterized by bulk heat transport  $Nu \sim Ra^{1/2}$ with logarithmic corrections \cite{Kraichnan62,Chavanne97} is not, in fact, supported by their data.

\vspace{-0.40 cm}

\begin{acknowledgments}
\vspace{-0.40 cm}
This work was supported by National Science Foundation Grants DMS-1515161 and DMS-1813003, Vetenskapsr\r{a}det No. 638-2013-9243, and All Souls College, Oxford.
\end{acknowledgments}


\begin{thebibliography} {100}

\bibitem{zhu2018}{X. Zhu, V. Mathai, R.J.A.M. Stevens, R. Verzicco, and D. Lohse, Phys. Rev. Lett. {\bf 120}, 144502 (2018). The $(Ra,Nu)$ data are tabulated in the Supplemental Material.}

\bibitem{Kraichnan62}{R. Kraichnan, Phys. Fluids {\bf 5}, 1374 (1962).}

\bibitem{Chavanne97}{X. Chavanne, F. Chill\`a, B. Castaing, B. H\'ebral, B. Chabaud and J. Chaussy, Phys. Rev. Lett. {\bf 79}, 3648 (1997).}

\bibitem{LPK2001}{L.~P. Kadanoff, Phys. Today {\bf  54}, 34 (2001).}

\bibitem{AGL}{G. Ahlers, S. Grossmann and D. Lohse, Rev. Mod. Phys. {\bf 81}, 503 (2009).}

\bibitem{DTW}{C. R. Doering, S. Toppaladoddi and J. S. Wettlaufer, Phys. Rev. Lett. {\bf 123}, 259401 (2019).}

\bibitem{priestley}{C.~H.~B. Priestley, Aust. J. Phys. {\bf 7}, 176 (1954).}

\bibitem{malkus}{W.~V.~R. Malkus, Proc. R. Soc. A {\bf 225}, 196 (1954).}

\bibitem{howard}{L.~N. Howard, in Applied Mechanics, Proceedings of the Eleventh International Congress of Applied Mechanics, Munich, Germany, ed. H. G\"{o}rtler (Springer, Berlin, 1966), 1109-1115.}

\bibitem{powerlaw}{M.P. Stumpf and M.A. Porter, Science {\bf 335}(6069), 665 (2012).}

\bibitem{ZhuReply}{X. Zhu, V. Mathai, R.J.A.M. Stevens, R. Verzicco, and D. Lohse, Phys. Rev. Lett. {\bf 123}, 259402 (2019).}

\bibitem{remark}{Two additional data $(Ra, Nu) = (2.126 \times 10^{14}, 2032)$ and $(4.522 \times 10^{14}, 2647)$ were extracted from \cite{ZhuReply}.  Physical Review Letters prohibited reference to these additional two points in the analysis published in \cite{DTW}.}

\bibitem{charlie2009}{H. Johnston and C.~R. Doering, Phys. Rev. Lett. {\bf 102}, 064501 (2009).}

\bibitem{VC}{R. Verzicco and R. Camussi, J. Fluid Mech. {\bf 477}, 19 (2003).}

\bibitem{SVL}{R.J.A.M. Stevens, R. Verzicco and D. Lohse, J. Fluid Mech. {\bf 643}, 495 (2010).}

\bibitem{urban1}{P. Urban, V. Musilova, and L. Skrbek, Phys. Rev. Lett. {\bf 107}, 014302 (2011).}

\bibitem{AGL_FIG.4}{See, e.g., data in Figure 4 and related text in \cite{AGL}.}

\bibitem{Sun}{C. Sun, L.-Y. Ren, H. Song and K.-Q. Xia,  J. Fluid Mech. {\bf 542}, 165-174 (2005).}

\end{thebibliography}
\end{document}